\documentclass{nature}
\usepackage{graphicx}
\usepackage{makecell}
\usepackage{subcaption}
\makeatletter
\let\saved@includegraphics\includegraphics
\AtBeginDocument{\let\includegraphics\saved@includegraphics}
\renewenvironment*{figure}{\@float{figure}}{\end@float}
\makeatother

\usepackage{lineno}
\usepackage[x11names]{xcolor}

\newcommand{\mn}{{Mon. Not. R. Astron. Soc.}}
\newcommand{\mnras}{\mn}
\newcommand{\aj}{{Astron. J.}}
\newcommand{\apj}{{Astrophys. J.}}
\newcommand{\apjl}{{Astrophys. J. Lett.}}
\newcommand{\apjs}{{Astrophys. J. Supp.}}
\newcommand{\apss}{{Astrophys. and Space Sc.}}

\newcommand{\aapr}{{Astron. Astrophys. Rev.}}
\newcommand{\aap}{{Astron. Astrophys.}}
\newcommand{\nat}{{Nat.}}
\newcommand{\pasj}{{Publ. Astron. Soc. Jpn }}

\newcommand{\pasp}{{Pub. Ast. Soc. Pac.}}

\newcommand{\ssr}{Space Science Reviews}

\title{A highly magnetised and rapidly rotating white dwarf as small as the Moon}

\author{Ilaria Caiazzo\footnote{email: ilariac@caltech.edu} $^1$, Kevin B. Burdge$^1$, James Fuller$^1$, Jeremy Heyl$^2$, S. R. Kulkarni$^1$, Thomas A. Prince$^1$, Harvey B. Richer$^2$, Josiah Schwab$^3$, Igor Andreoni$^1$, Eric C. Bellm$^4$, Andrew Drake$^1$, Dmitry A. Duev$^1$, Matthew J. Graham$^1$, George Helou$^5$, Ashish~A.~Mahabal$^{1,6}$ Frank J. Masci$^5$, Roger Smith$^7$, Maayane T. Soumagnac$^{8,9}$}

\begin{document}

\maketitle

\begin{affiliations}
 
 \item Division of Physics, Mathematics and Astronomy, California Institute of Technology, Pasadena, CA 91125, USA
 \item Department of Physics and Astronomy, University of British Columbia, Vancouver, BC, V6T1Z1, Canada
 \item Department of Astronomy and Astrophysics, University of California, Santa Cruz, CA 95064, USA
 \item Department of Astronomy, University of Washington, Seattle, WA 98195, USA
 \item IPAC, California Institute of Technology, 1200 E. California Blvd, Pasadena, CA 91125, USA
 \item Center for Data Driven Discovery, California Institute of Technology, Pasadena, CA 91125, USA
 \item Caltech Optical Observatories, California Institute of Technology, Pasadena, CA 91125, USA
 \item Lawrence Berkeley National Laboratory, 1 Cyclotron Road, Berkeley, CA 94720, USA
 \item Department of Particle Physics and Astrophysics, Weizmann Institute of Science, Rehovot 76100, Israel
 
\end{affiliations}

\begin{abstract}
White dwarfs represent the last stage of evolution of stars with mass less than about eight times that of the Sun and, like other stars, are often found in binaries\cite{ELMVIII,2020ApJ...905...32B}. If the orbital period of the binary is short enough, energy losses from gravitational-wave radiation can shrink the orbit until the two white dwarfs come into contact and merge\cite{2015ApJ...805L...6S}. Depending on the component masses, the merger can lead to a supernova of type Ia or result in a massive white dwarf\cite{2014MNRAS.438...14D}. In the latter case, the white dwarf remnant is expected to be highly magnetised\cite{2008MNRAS.387..897T,2012ApJ...749...25G} because of the strong magnetic dynamo that should arise during the merger, and be rapidly spinning from the conservation of the orbital angular momentum\cite{2021ApJ...906...53S}. Here we report observations of a white dwarf, ZTF J190132.9+145808.7, that exhibits these properties, but to an extreme: a rotation period of 6.94 minutes, a magnetic field ranging between 600 megagauss and 900 megagauss over its surface, and a stellar radius of $2,140^{+160}_{-230}$~km, slightly larger than the radius of the Moon. Such a small radius implies that the star’s mass is close to the maximum white-dwarf mass, or Chandrasekhar mass. ZTF J190132.9+145808.7 is likely to be cooling through the Urca processes (neutrino emission from electron capture on sodium) because of the high densities reached in its core. 
\end{abstract}

\begin{figure}
    \centering
    \includegraphics[width=0.8\textwidth]{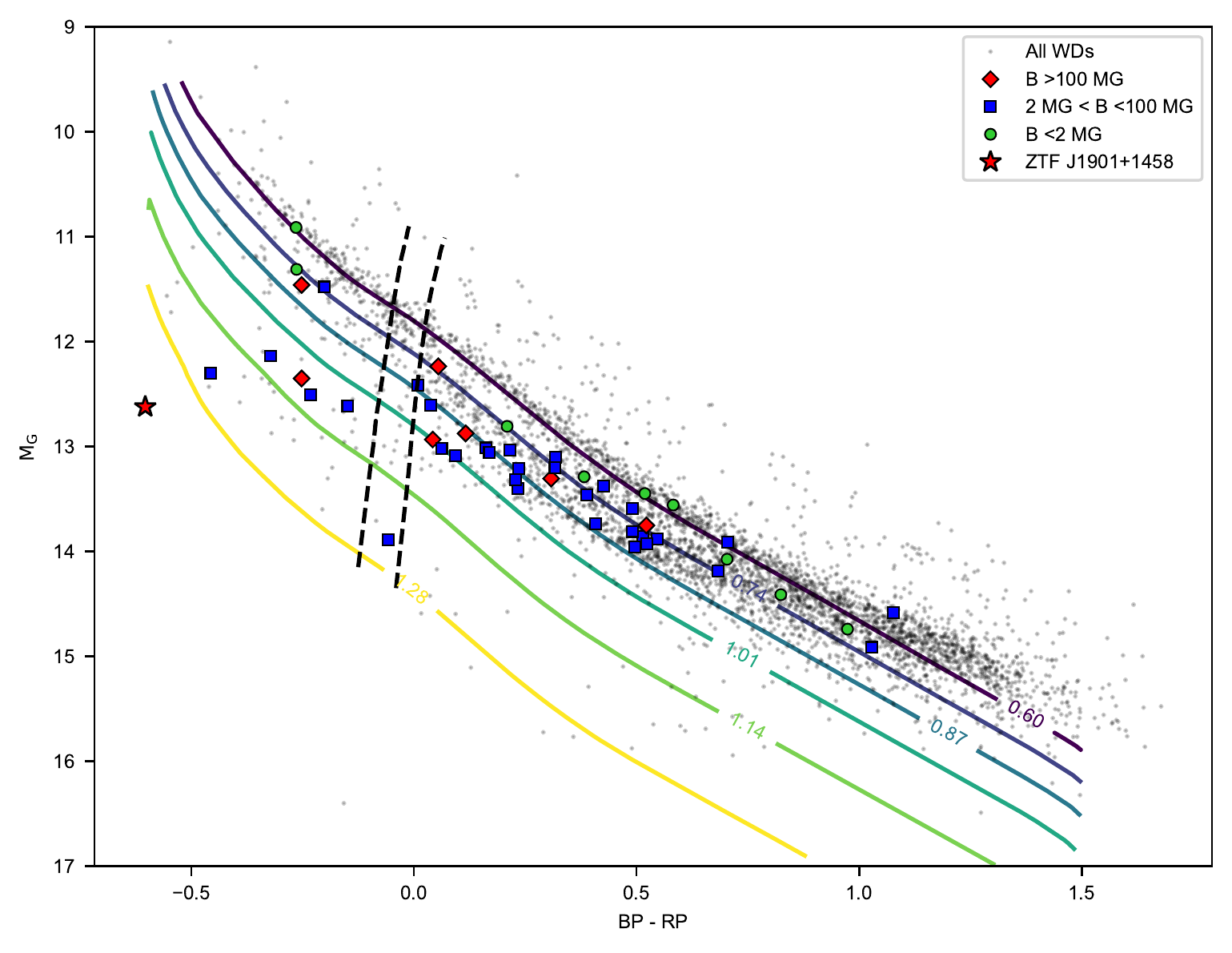}
    \caption{{\bf Gaia colour-magnitude diagram.} Gaia color-magnitude diagram for the white dwarfs that are within 100~pc from Earth and within the SDSS footprint\cite{2020ApJ...898...84K}, where the x-axis depicts the difference between the Gaia BP and RP bands, and the y-axis the absolute magnitude in the Gaia G filter. Solid lines show theoretical cooling tracks for white dwarfs with masses between 0.6~M$_\odot$ (top) and 1.28~M$_\odot$ (bottom), equally spaced in mass; the atmosphere is assumed to be hydrogen-dominated\cite{2006AJ....132.1221H} and the interior composition to be carbon-oxygen\cite{2001PASP..113..409F} for $M<1.1$~M$_\odot$ and oxygen-neon\cite{2019A&A...625A..87C} for $M>1.1$~M$_\odot$. Coloured markers indicate white dwarfs for which a magnetic field was measured\cite{2015SSRv..191..111F}. ZTF~J1901+1458 is shown as a red star, and its location in the colour-magnitude diagram reveals its high mass. Vertical dashed lines indicate the location of the ZZ-Ceti instability strip\cite{2012A&A...539A..87V} (the pre-white dwarf, or DOV, instability strip lies above the plot\cite{2014PASJ...66...76M,2012ApJ...755..128Q}) .
    Reddening corrections were applied only to ZTF~J1901+1458; as the objects in the sample are close, reddening is expected to be small. 1$\sigma$ error bars are smaller than the size of the coloured markers,and are omitted for the black background dots for clarity.}
    \label{fig:cmd}
\end{figure}

\begin{figure}
    \centering
    \includegraphics[width=\textwidth]{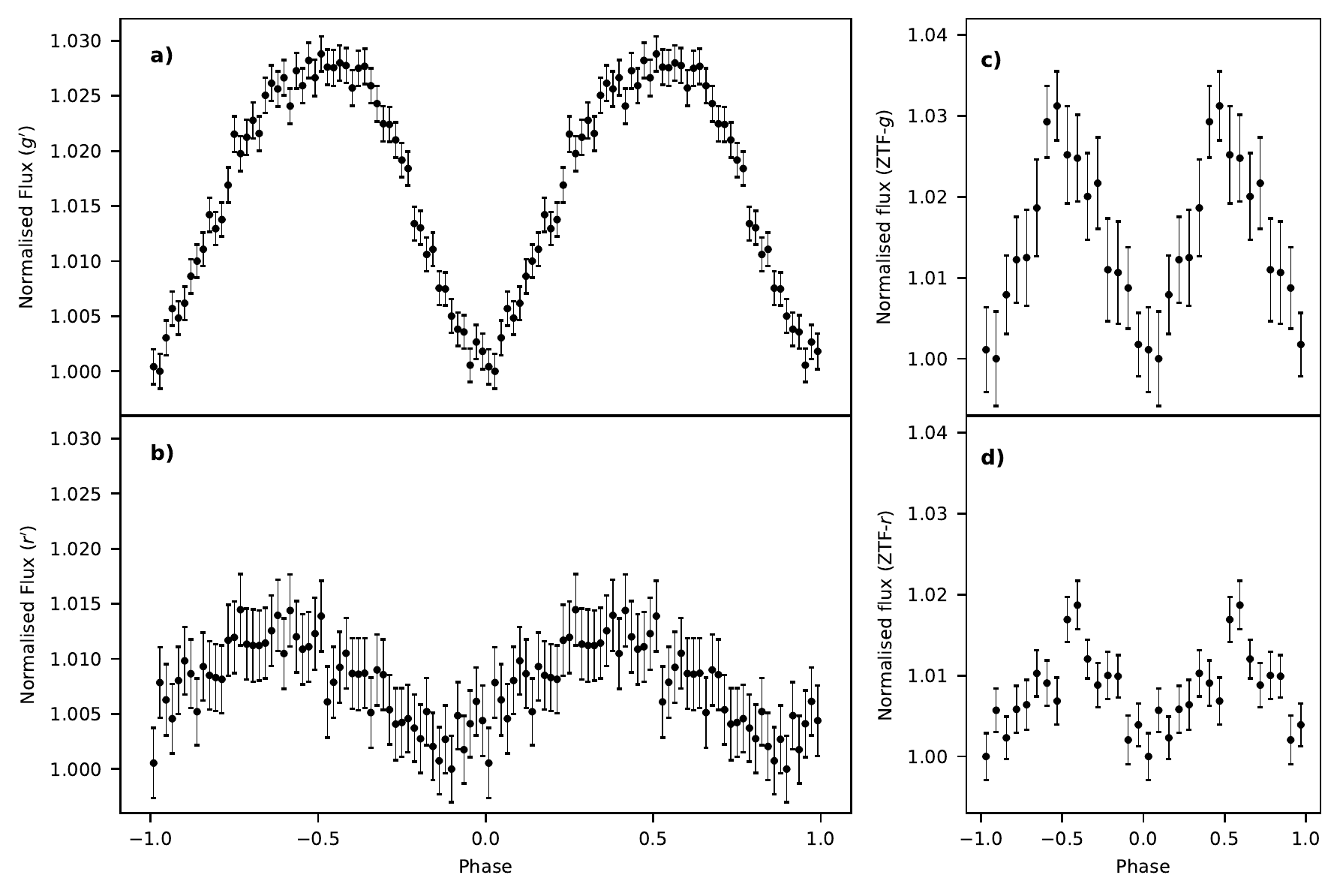}
    \caption{{\bf ZTF~J1901+1458 lightcurve} The left panels show the binned CHIMERA lightcurve phase-folded at a period of 6.94 minutes in the $g^\prime$-band (\textbf{a}) and in the $r^\prime$-band (\textbf{b}). The flux has been normalised to the minimum of the lightcurve in each band. The amplitude of the photometric variation is higher in the $g^\prime$-band (about 3\% peak-to-peak) than in the $r^\prime$-band ($\sim1.5$\%). Additionally, the two filters show a difference in phase: the red lags the green by about 51 s.  The right panels show the similarly normalised ZTF discovery lightcurve in the ZTF $g$-band (\textbf{c}) and $r-$band (\textbf{d}). The error bars indicate 1$\sigma$ errors.} 
    \label{fig:lc}
\end{figure}

Using the The Zwicky Transient Facility\cite{2019PASP..131a8002B} (ZTF) we searched for short period objects that lie below the main white dwarf cooling sequence in the Gaia\cite{2016A&A...595A...1G} colour-magnitude diagram (see Fig.~\ref{fig:cmd}).
ZTF~J190132.9+145808.7 (hereafter ZTF~J1901+1458) 
showed promising small photometric variations.  Followup with \emph{CHIMERA}\cite{2016MNRAS.457.3036H}, a high-speed imaging photometer on the 200-inch Hale telescope, confirmed a period of 6.94 minutes (see Fig.~\ref{fig:lc}). The period of ZTF~J1901+1458 is unusually short for a white dwarf, as  white dwarf rotational periods typically are upwards of hours\cite{2017ApJS..232...23H}.
The period and ephemeris are listed in Table~\ref{tab:par}.

\begin{figure}
    \centering
    \includegraphics[width=\textwidth]{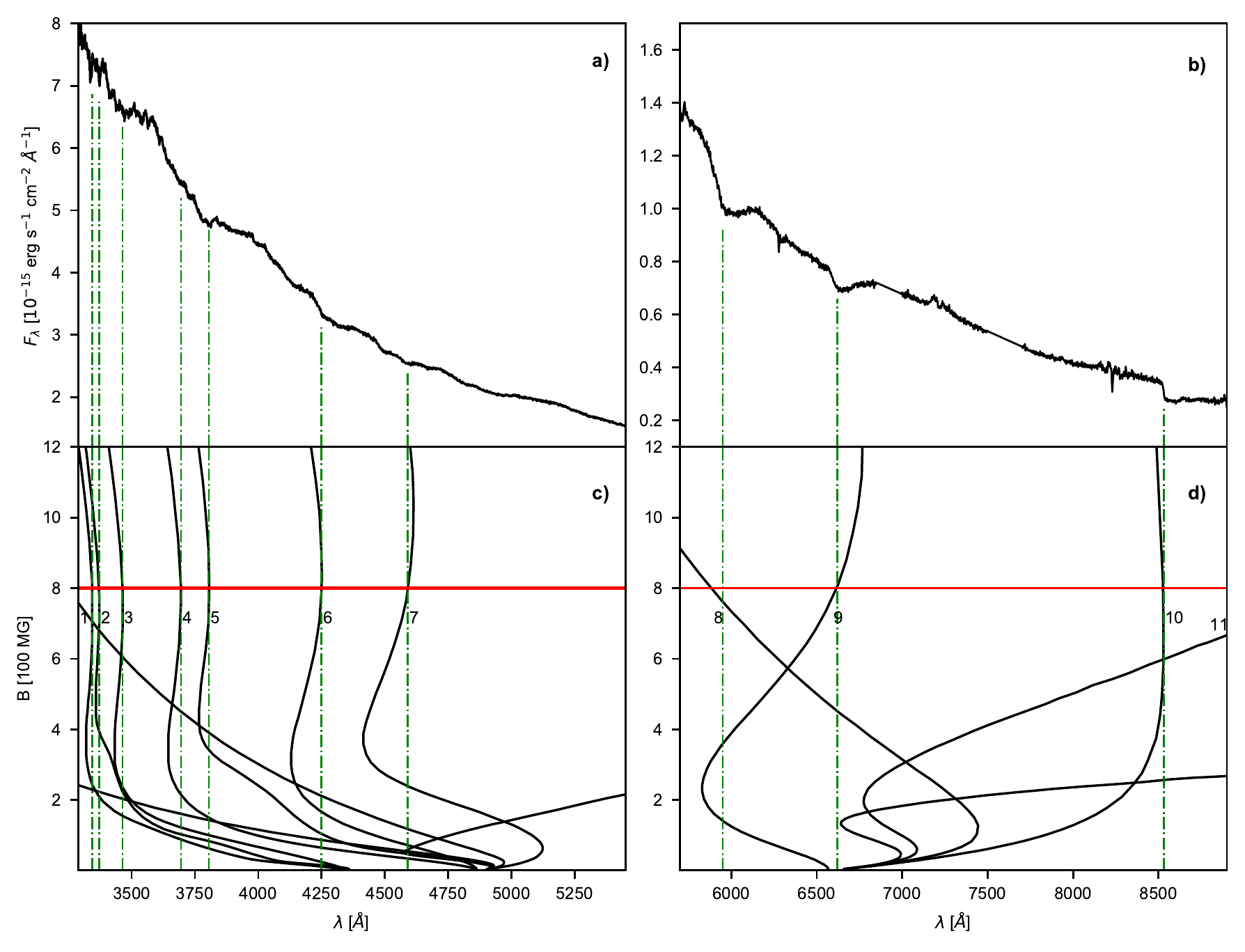}
    \caption{{\bf ZTF~J1901+1458 optical spectrum.} The LRIS phase-averaged spectrum of ZTF~J1901+1458 is shown in black in the upper panels for the blue (\textbf{a}) and red (\textbf{b}) sections. In (\textbf{c)}-(\textbf{d})  the spectrum is compared to predicted line positions of H$\alpha$, H$\beta$ and H$\gamma$ as a function of magnetic field\cite{1994asmf.book.....R}, showing that the white dwarf is characterised by a field strength of about 800 MG (red horizontal solid line). The identified absorption features are highlighted by the green dash-and-dotted lines and numbered, and the respective transitions are listed in Extended Data Table~\ref{tab:transitions}. } 
    \label{fig:sp}
\end{figure}

We undertook phase-resolved spectroscopy using the Low Resolution Imaging Spectrometer (\emph{LRIS})\cite{1995PASP..107..375O} on the 10-m W. M. Keck I Telescope. As can be seen from Fig.~\ref{fig:sp}, the phase-averaged spectrum exhibits broad and shallow features that we identify as hydrogen absorption lines in a high magnetic field. 
The presence of a strong magnetic field results in splitting and proportional shifting of the zero-field energy levels, leading to line broadening.  To identify the field strength, we considered all the allowed bound-bound hydrogen transitions (tabulated in\cite{1994asmf.book.....R}) and, as shown in Fig.~\ref{fig:sp}, we find a satisfactory explanation for the spectrum. The identified spectral lines are listed in Extended Data Table~\ref{tab:transitions}.

We find that most of the spectral features are well characterised by a magnetic field strength of 800 million Gauss (MG, red horizontal line in Fig.~\ref{fig:sp}), comparable to the field detected on the most magnetic white dwarfs known\cite{2015SSRv..191..111F}. As the absorption features indicate an average field strength over the surface, the field at the magnetic pole is bound to be higher. From the phase-resolved spectra (Extended Data Fig.~\ref{fig:pr1} and~\ref{fig:pr2}) we see that some of the features become narrower or broader depending on the phase, and the feature at $\sim4600$~\AA\ shifts in wavelength, which accounts for the dip at $\sim4500$~\AA\ in the co-added spectrum. This means that in some regions of the surface the magnetic field is as low as 600~MG. 
The explanation for the photometric variation, confirmed by the variations of absorption features with phase, 
is thus the combination of magnetic dichroism and rotation: the high magnetic field causes variations in the continuum opacities and in the surface temperature, and therefore, as the star rotates, we detect changes in flux as a function of the field strength across the stellar surface. Depending on the magnetic field configuration, this dichroism can account for up to  10\% photometric variation\cite{1997MNRAS.292..205F}, so it can easily account for the 3\% amplitude observed in ZTF~J1901+1458. ZTF~J1901+1458's period could also be consistent with non-radial pulsations; however, its temperature and surface gravity place it far away from theoretical predictions for known instabilities
(see Fig.~\ref{fig:cmd}), and its magnetic field may be strong enough to suppress gravity-mode pulsations (see the Method section for further discussion).

To determine the temperature and radius of the white dwarf, we obtained photometric measurements in the near UV using the UVOT (Ultraviolet/Optical Telescope) instrument\cite{2005SSRv..120...95R} on the Neil Gehrels Swift Observatory\cite{2004ApJ...611.1005G} while the Pan-STARRS Survey\cite{2016arXiv161205560C} (PS1) and 
the Gaia mission provided optical photometry and parallax, respectively.  We estimated the temperature, reddening and radius from the photometry, by comparing the observations with non-magnetic white dwarf atmosphere models. Because the white dwarf is very hot, the photometric constraints on the temperature are weak; however, the precise distance measurement from Gaia allows us to obtain a good estimate for the radius. We found the effective temperature, stellar radius and interstellar reddening to be  $T_{\rm{eff}}=46,000^{+19,000}_{-8,000}$~K, $R_*=2,140^{+160}_{-230}$~km and $E(B-V)=0.044^{+0.017}_{-0.015}$, respectively (see the Methods section for more details).  The radius is smaller than those measured for other white dwarfs and only slightly larger than that of the Moon. As explained below, the small radius also means that ZTF J1901+1458 maybe the most massive white dwarf yet discovered.

\begin{figure}
    \centering
    \includegraphics[width=0.9\textwidth]{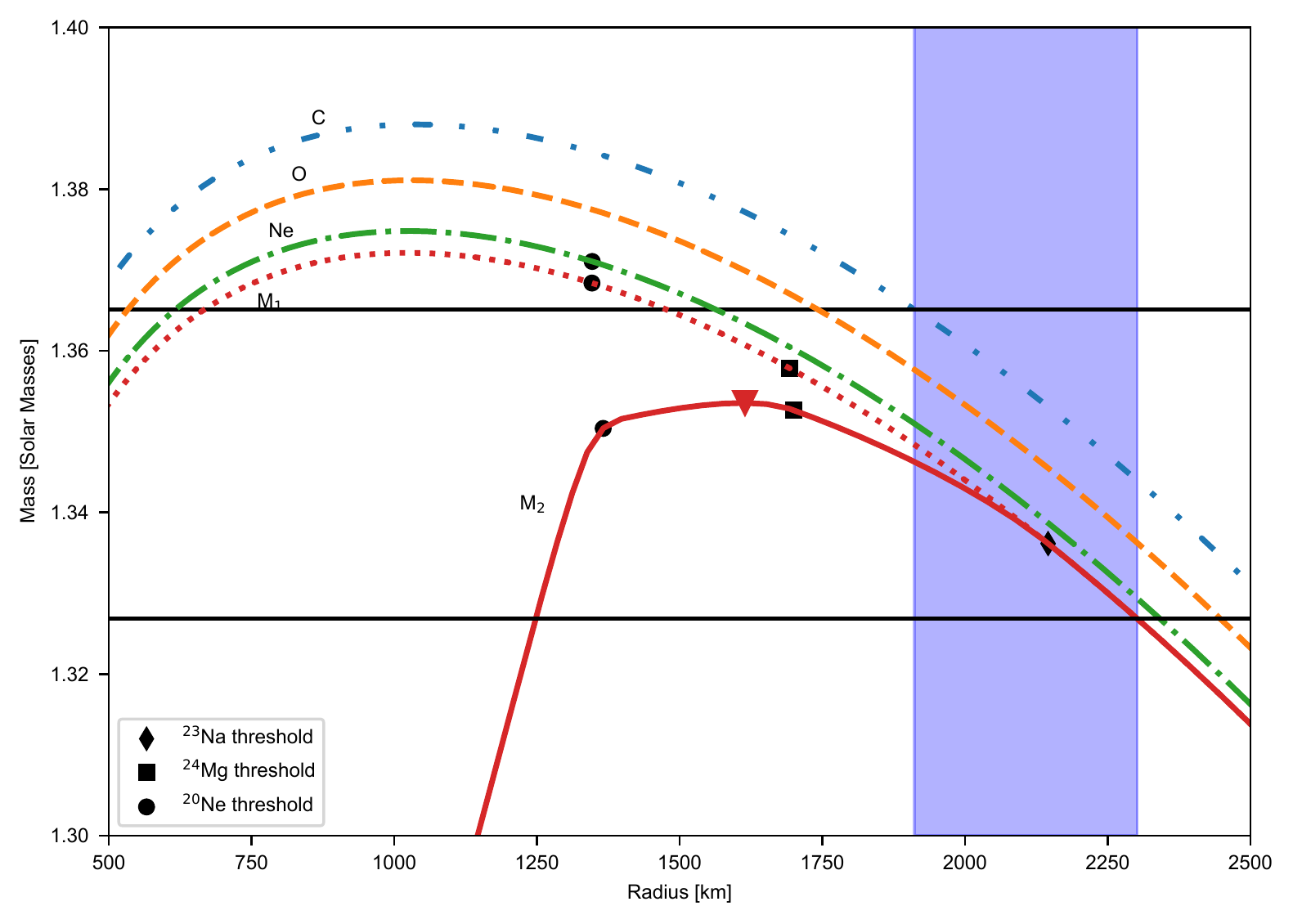}
    \caption{{\bf Mass-radius relation.} The white dwarf mass-radius relation at the temperature of ZTF~J1901+1458 ($\sim$46,000~K) is shown for different homogeneous compositions neglecting the effect of electron captures: carbon C; oxygen O; neon Ne; and the mixture that is thought to result from carbon burning (M$_1$), specifically 58\% oxygen, 30\% neon, 2\% carbon, 5\% sodium, and 5\% magnesium\cite{2019A&A...625A..87C}. 
    The lowermost red solid curve (M$_2$) traces the result from the carbon-burning-ash mixture including the effect of electron captures (see text). The black marks along each track indicate the radius at which the central density reaches the threshold for inverse beta-decay of neon (circle), magnesium (square) and sodium (diamond).  The vertical blue-shaded area indicates the observed radius of ZTF~J1901+1458 within 1$\sigma$ errors. The red triangle indicates the maximum mass of 1.354~$M_\odot$ at a radius of 1,600~km along the track of the mixed composition. The horizontal black solid lines highlight the mass determined using a radius of 1,910 km for the upper line (pure C),  1.365 M$_\odot$ and of 2,300 km for the lower line (carbon-burning-ash mixture),  1.327 M$_\odot$.
    }
    \label{fig:mr}
\end{figure}

The mass can be inferred from the mass-radius relation which, as can be seen in Fig.~\ref{fig:mr}, is composition dependent.  White dwarf descendants of single stars with mass above $1.1$ solar masses (M$_{\odot}$) are expected to be mostly oxygen and neon, with traces of carbon, sodium and magnesium\cite{2007A&A...476..893S,2019A&A...625A..87C}. 
Even if born from a merger, compressional heating due to rapid accretion is expected to ignite off-center carbon burning\cite{2012ApJ...748...35S}, resulting in an O/Ne white dwarf\cite{2021ApJ...906...53S}. We conclude that, depending on the composition, ZTF\,J1901+1458 has a mass between 1.327 and 1.365 solar masses.

Two other extremely massive white dwarfs for which a radius has been measured are RE~J0317-853\cite{2010A&A...524A..36K} and WD~1832+089\cite{2020MNRAS.499L..21P}, both with a radius of about 2,500~km. Curiously, they are both variable with short periods (6 and 12 minutes, respectively). While WD~1832+089 does not show evidence for magnetism, RE~J0317-853 appears to have a field that varies between 185 and 425~MG over its surface\cite{1997MNRAS.292..205F}.
Thus both ZTF~J1901+1458 and RE~J0317-853, with their rapid rotation, high mass and magnetism, are likely remnants of a white dwarf merger. 

At the densities reached in the centre of ZTF~J1901+1458, the nuclei of some elements may undergo electron capture (also called inverse beta-decay), removing electrons that contribute to the degeneracy pressure that keeps the star from collapsing.  
This lowers the maximum mass that can be sustained against gravity and reduces the equilibrium radius for a fixed mass (see Fig.~\ref{fig:mr}, solid red curve). If ZTF~J1901+1458 has an oxygen-neon internal composition (as is expected from its mass\cite{2019A&A...625A..87C,2021ApJ...906...53S}), its central density is right at the threshold for electron capture on $^{23}$Na, and its mass is within 2\% of the highest possible mass for a white dwarf.

Neutrinos produced by inverse beta-decay escape and carry away energy, contributing to the cooling of the white dwarf.  Based on the  luminosity of the white dwarf\cite{2019A&A...625A..87C}, we estimate the temperature in the core to be about $2-3\times 10^7$~K. At such a high central temperature and density, the neutrino cooling of ZTF~J1901+1458 will dominated by the "Urca" process\cite{1970Ap&SS...7..374T,2017MNRAS.472.3390S} acting on $^{23}$Na.  This unusual neutrino cooling makes an age determination difficult. A recent work\cite{Schwab2021} studied the evolution of Urca-cooling white dwarfs; from the same models, we estimate the cooling age of ZTF~J1901+1458 to be between 10 and 100~Myr.

The mass-radius relation in Fig.~\ref{fig:mr} was calculated assuming that the core composition is homogeneous -- a good assumption since ZTF~J1901+1458 is less than 100~Myr old. However, over a few hundred million years, the heaviest elements, including Na, will gradually sink to the centre\cite{2002ApJ...580.1077D}.
If the star lies at the small end of the radius constraint and if at least sixty percent of the $^{23}$Na manages to sink and undergo beta decay before the core crystallises and sedimentation stops, electron-capture on $^{24}$Mg would ensue. The star would shrink and the internal pressure would no longer be able to support the star, as the maximum allowed mass for the new composition would be lower than the mass of the white dwarf (see the Methods section for a more detailed discussion). The star would therefore collapse and heat up, leading to the onset of electron capture onto Ne and to the ignition of oxygen nuclear burning. The white dwarf would then undergo a disruptive thermonuclear supernova or implode to form a neutron star\cite{2017MNRAS.472.3390S}.

The probability of collapse is highly uncertain, as it depends on the timescales of sedimentation and crystallisation, both unconstrained at these high masses. However, the possibility of this novel formation channel for neutron stars is intriguing. If, upon collapse, no angular momentum is lost and magnetic flux is conserved, the newly born neutron star, with a magnetic field strength of $\sim2\times10^{13}$~G and a spin period of $\sim15$~ms, would resemble a young pulsar. Owing to the gradual collapse, the neutron star would likely not receive a strong natal velocity kick. We would therefore expect such a neutron star population to be more concentrated to the Galactic plane. Furthermore, the proximity of ZTF~J1901+1458 (41\,parsecs) means it is not a rare type of object, and thus this formation channel would contribute appreciably to the total neutron star population. ZTF is currently discovering large numbers of similarly massive and rapidly variable white dwarfs. This
enlarged sample will help us better understand the origin and fate of such objects.

\begin{table}
\centering
 \caption{\textbf{ZTF~J1901+1458 parameters}. The ephemeris $T_0$, Barycentric Modified Julian Date in barycentric dynamical time (BMJD$_{\rm{TBD}}$), corresponds to a minimum in the light curve and it is given separately for the ZTF $g$-band and $r$-band because the latter lags behind the former by 51 seconds. Here, $T_0$ corresponds to the minimum of each  lightcurve.}
\medskip
\begin{tabular}{cccc}
\hline
\makecell{$\rm{Gaia~ID}$} & \makecell{$\rm{Parallax~[mas]}$} &\makecell{$\mu_{\rm{RA}}~\rm{[mas/yr]}$} & \makecell{$\mu_{\rm{DEC}}~\rm{[mas/yr]}$}\\ 
$4506869128279648512$ & $24.13\pm0.06$ & $95.50\pm0.09$ & $72.45\pm0.09$ \\
\hline
\makecell{$T_{\rm{eff}}~\rm{[K]}$} & \makecell{$R_*~\rm{[km]}$}& \makecell{$M~[\rm{M}_\odot]$} & \makecell{$E(B-V)$} \\
$46,000^{+19,000}_{-8,000}$ & $2,140^{+160}_{-230}$ & $1.327-1.365$ & $0.044^{+0.017}_{-0.015}$ \\ 
\hline
\makecell{$T_0$ ($g$) $[\rm{BMJD}_{\rm{TDB}}]$} & \makecell{$T_0$ ($r$) $[\rm{BMJD}_{\rm{TDB}}]$} & \makecell{$P~\rm{[s]}$} & \makecell{$\dot{P}~[\rm{s~s}^{-1}]$} \\
$59079.217290\pm0.000012$ & $59079.21670\pm0.00006$ & $416.2427\pm0.0002$ & $<10^{-11}$ \\
\hline
\end{tabular}
\label{tab:par}
\end{table}

\clearpage

\begin{methods}

\renewcommand{\tablename}{Extended Data Table}
\setcounter{table}{0} 
\renewcommand{\figurename}{Extended Data Figure}
\setcounter{figure}{0} 

\subsection{Period Detection}
Our search for periodicity in massive white dwarfs was part of a broader search for periodic variability on and around the white dwarf cooling track with ZTF, which has already yielded several results, including finding numerous double white dwarf binaries\cite{Burdge2019a,Burdge2019b,2020ApJ...905...32B}. The targets were selected using Pan-STARRS (PS1) source catalogue\cite{2016arXiv161205560C}, cross-matched with a white dwarf catalogue\cite{2019MNRAS.482.4570G}, after imposing a photometric colour selection of $(g-r)<0.2$ and $(r-i)<0.2$. As the sensitivity of period finding depends strongly on the number of samples in the lightcurve, we limited the search to those targets for which 50 or more photometric $5 \sigma$ detections are available in the ZTF archival data. In order to maximise the number of epochs for each lightcurve, we combined data from multiple filters by computing the median magnitude in each filter, and shifting the $g$- and $i$-band so that their median magnitude matched the $r$-band data. We used a graphics processing unit (GPU) implementation of the conditional entropy period finding algorithm\cite{Graham2013}. We cross-matched our candidates with the Gaia DR2 catalogue\cite{2018A&A...616A...1G} and visually inspected the lightcurves of those objects that lie below the main white dwarf cooling track in the Gaia colour-magnitude diagram. ZTF~J1901+1458 stood out because of the high-significance detection of its short period and its blue and faint location in the colour-magnitude diagram (see Fig.~\ref{fig:cmd}).

\subsection{Magnetic Field}
At low magnetic field strengths, the effect of magnetic field on hydrogen transitions can be calculated as a perturbation to the zero-field electron wavefunction. The perturbation lifts the degeneracy in the $m$ quantum number, and each transition is split into three Zeeman components, corresponding to a change in $m$ of $+1$, 0 and $-1$. In the strong regime (above $\sim100$~MG), the magnetic and Coulomb terms are comparable, and the wavefunction does not have a spherical symmetry anymore. The perturbation method is not viable in this case, and the energies and oscillator strengths of hydrogen transitions in this regime were calculated using variational methods\cite{1984JPhB...17...29R,1984JPhB...17.1301F,1985PASP...97..333H,1985A&A...149..102W}. An important characteristic of high-field transitions is that some transitions become `stationary', i.e. they go through a minimum or a maximum in energy and, at those field strengths, appreciable changes in $B$ only yield small changes in wavelength. The magnetic broadening at these transitions is considerably reduced, and therefore the absorption features are still detectable even after averaging the field over the surface of the white dwarf. We considered all the bound-bound transitions tabulated in Ruder et al. 1994\cite{1994asmf.book.....R} that follow selection rules, and we found that the features in the spectrum of ZTF~J1901+1458 correspond to stationary lines in a $\sim800$~MG field. The wavelengths of the transitions in this regime as a function of magnetic field strength are shown in the lower panels of Fig.~\ref{fig:sp}. The identified features are numbered and the corresponding transitions are listed in Extended Data Table~\ref{tab:transitions}, where the states are labelled by their asymptotic quantum numbers $nlm$ for the zero-field states and by the numbers $NM\mu$ for $B\rightarrow\infty$, where $\mu$ defines the $z$ parity of the states $[\pi = (-1)^\mu]$. A few lines show variations with phase (see Extended Data Fig~\ref{fig:pr1} and~\ref{fig:pr2}), which can be attributed to the variation of the magnetic field strength across the surface. In particular, some of the features become narrower or broader at different times, indicating that the field is more or less uniform over the surface at different rotation phases. Additionally, the feature at $\sim4600$~\AA (line 7) shifts in wavelength, going as low as $\sim4500$~\AA, which indicates that the magnetic field is as low as 600~MG on some regions on the surface of the white dwarf. This is also confirmed by the contamination of line 10 by line 11.

\subsection{Photometric Fitting}
To determine the radius and effective temperature of ZTF~J1901+1458, we made use of the available Pan-STARRS\cite{2016arXiv161205560C} photometry and the Gaia\cite{2016A&A...595A...1G,2018A&A...616A...1G} parallax. In addition, we obtained Swift\cite{2004ApJ...611.1005G} UVOT\cite{2005SSRv..120...95R} photometry (TOO proposal number 14380, target ID 13650). The photometric data used in the fitting is listed in Extended Data Table~\ref{tab:photometry}. As the errors in the Pan-STARRS photometry are lower than the photometric variation, we used an error of $0.02$ magnitudes instead to take into account the error induced by variability. In order to estimate the temperature and reddening, we fitted the photometric data with the white dwarf 1-D model DA (hydrogen dominated) atmospheres of Tremblay et al. 2011\cite{2011ApJ...730..128T}. The value of the radius is then estimated thanks to the parallax measurement from Gaia. In order to account for extinction, we applied reddening corrections to the synthetic spectra (and polynomials) using the Cardelli et al. 1989\cite{1989ApJ...345..245C} extinction curves, available at \texttt{https://www.stsci.edu/}. From the corrected models, we computed synthetic photometry using the \texttt{pyphot} package (\texttt{https://mfouesneau.github.io/docs/pyphot/}).


For the fit, we used a Levenberg-Marquardt algorithm, and the free parameters were the effective temperature $T_{\rm{eff}}$ and radius $R_*$ of the white dwarf as well as the interstellar reddening $E(B-V)$. For the reddening, we imposed a prior based on the distribution of interstellar extinction measured by Gaia for the stars in the same area of the sky as ZTF J1901+1458. In particular, we selected the stars that are within 5 degrees of the location of white dwarf and within 50 pc from Earth (see Extended Data Figure~\ref{fig:ebmv}).

The best fit is shown in Extended Data Fig.~\ref{fig:phot}, while the left panel of Extended Data Fig.~\ref{fig:corner} illustrates the corner plots. In the fit, we assumed the nominal value of the Gaia parallax for the distance. As a change in distance would have the effect of changing the normalisation in the spectrum, the error in the distance only influences the error in the stellar radius, and therefore we included the error in the distance in the error in the radius. The errors quoted are the statistical errors derived from the Monte Carlo simulation shown in Extended Data Fig.~\ref{fig:corner}. They do not include possible modelling errors due to the effect of the magnetic field; however, as the photometric variation is quite small, we do not expect the magnetic field to introduce a significant error on the synthetic values.

In order to verify whether our estimates are robust against the uncertainties in the spectral models, we employed a different set of models for hydrogen dominated atmospheres, the ones developed by Bohlin et al 2020\cite{2020AJ....160...21B}, and performed the same analysis introduced above. The results are shown in the right panel of Extended Data Fig.~\ref{fig:corner}: the estimates of temperature, radius and reddening all agree within the 1$\sigma$ errors. Magnetic models that tackle the effects of magnetic field on the continuum opacities are currently unavailable, and therefore we cannot estimate the errors due to neglecting the field.

\subsection{Theoretical Mass-Radius Relation}
In order to derive the mass of the white dwarf, we computed the mass-radius relation employing an equation-of-state that includes Coulomb corrections to the pressure and energy of a degenerate Fermi gas\cite{2000ApJS..126..501T,pyt-helm}, as done in Hamada and Salpeter 1961\cite{1961ApJ...134..683H} (H\&S). Since ZTF~J1901+1458 is so compact, general relativistic corrections are important and therefore, contrary to H\&S, we did not integrate the Newtonian hydrostatic equilibrium equation, but rather the Tolman-Oppenheimer-Volkoff (TOV) equation\cite{1939PhRv...55..364T,1939PhRv...55..374O}. The results are shown in Fig.~\ref{fig:mr} for several compositions. Our calculations were performed assuming the zero-temperature limit because the temperature of the white dwarf is insufficient to change its structure significantly.  For the composition of the carbon-burning ash, we use the results from Camisassa et al.\cite{2019A&A...625A..87C}. For the density thresholds for the electron capture onto magnesium and neon\cite{1980PASJ...32..303M}, we use the values of $4\times 10^9~\textrm{g cm}^{-3}$ and  $9\times 10^9~\textrm{g cm}^{-3}$ respectively. For sodium\cite{2017MNRAS.472.3390S} we use $1.7\times 10^9~\textrm{g cm}^{-3}$.  We estimate the temperature at a density of $1.7\times 10^9~\textrm{g cm}^{-3}$ to be about $3\times 10^7$~K; therefore, the neutrino cooling ZTF~J1901+1458, unlike for all other known white dwarfs, is cooling dominated by the Urca process on $^{23}$Na\cite{1970Ap&SS...7..374T,2017MNRAS.472.3390S} which makes an age determination difficult. A recent work\cite{Schwab2021} simulated the evolution of Urca-cooling white dwarfs with the Modules for Experiments in Stellar Astrophysics code (MESA)\cite{Paxton2011, Paxton2013, Paxton2015, Paxton2018, Paxton2019}, showing that the Urca processes are the main cooling mechanism in the core of stars like ZTF~J1901+1458 when they are younger than about 30 Myr. From the same models, we can estimate the cooling age of the white dwarf to be between 10 and 100~Myr.

\subsection{Sedimentation and collapse}
The solid red curve in Fig.~\ref{fig:mr} was calculated assuming that the core composition of the white dwarf is homogeneous and that all the sodium currently at densities above the threshold for electron capture has already undergone inverse beta decay, and similarly for magnesium and neon. As the sedimentation proceeds, much of the remaining $^{23}$Na will sink to densities above the threshold and undergo beta-decay, reducing the number of electrons in the star and reducing its radius. This means that the solid red curve in Fig.~\ref{fig:mr} will be lowered, because the equilibrium radius for any given mass will be smaller, and therefore the red triangle (the maximum mass allowed) will sink. This process can only be stopped if the core crystallises before enough sodium can reach the centre, as crystallisation would de facto freeze the composition gradient. If the star lies at the small end of the radius constraint and if at least sixty percent of the $^{23}$Na manages to sink and decay before the core crystallises, electron-capture on $^{24}$Mg would ensue and the radius of the white dwarf would shrink to about 1,550~km. The central density at this point would be $6 \times 10^{9}~\textrm{g cm}^{-3}$, still below the threshold for electron capture on neon, but the mass of the white dwarf would be above the maximum mass (the red triangle in Fig.~\ref{fig:mr} would have sunk below the current white dwarf mass). The internal pressure would be then insufficient to support the star, and the star would begin to collapse, heat up and start electron capture onto neon and nuclear burning of oxygen.

The possibility of collapse is highly uncertain as it depends on the timescales for crystallisation and for the sedimentation of sodium, both currently poorly constrained, especially for such an extremely massive white dwarf.

\subsection{Origin of Photometric Variability}
The photometric variability of ZTF J1901+1458 may arise from either rotation or pulsations, but we believe rotational modulation is more likely. A large fraction of hot DQs (hot white dwarfs with carbon-dominated atmospheres) are magnetic and photometrically variable, often with periods of several minutes\cite{Williams:16,Ferrario:20}. Although ZTF J1901+1458 is not a DQ white dwarf, its other properties such as high mass, high temperature, high magnetic field, and rotational modulation are similar to the hot DQ white dwarfs, albeit more extreme. A useful case study is that of the white dwarf SDSS J142625.71+575218.3, which varies with a period of $P\simeq 418$ seconds\cite{Montgomery:08} and has a magnetic field strength of $B \simeq 1.2$ MG\cite{Dufour:08}. Non-radial g mode pulsations were originally thought to be responsible for the photometric modulation of SDSS J1426. Follow-up observations revealed a tentative additional periodicity at 318 seconds\cite{Green:09}, but additional data ruled out the existence of this periodicity and set stringent limits on additional periodicities in SDSS 1426 and other variable DQ white dwarfs\cite{Williams:16}. This led Williams\cite{Williams:16} to conclude that rotational modulation is the most likely source of variability in SDSS 1426 and other hot DQ white dwarfs. Presently, there are no known strongly magnetic pulsating WDs, though some magnetic WDs could be pulsating below detectable levels.

Furthermore, the inferred temperature and surface gravity of ZTF J1901+1458 characterise it as an unlikely pulsator for known instability mechanisms. The temperature of ZTF J1901+1458, $\sim$50,000~K, is much higher than the predicted blue edge of the ZZ-Ceti instability strip, located at a temperature of about 12,500-14,000~K\cite{2012A&A...539A..87V,corsico:19}, and of the helium white dwarfs (DBV) instability strip, at about 30,000~K\cite{1999ApJ...516..887B} (the detection of hydrogen also discourages the DBV interpretation).  The DOV or GW Vir instability strip includes similar and higher temperatures than what we found for ZTF J1901+1458, but both instability mechanisms involved, the $\kappa-\gamma$ mechanism for carbon and oxygen and the $\epsilon-$mechanism for hydrogen, are inefficient at such high surface gravities (see for example Fig. 8 of Quirion et al 2012\cite{2012ApJ...755..128Q} and Fig. 6 of Maeda \& Shibahashi 2014\cite{2014PASJ...66...76M}). It is also possible that  magnetic fields suppress gravity mode pulsations by converting gravity waves to Alfv\'en waves\cite{fuller:15,lecoanet:17,loi:18,loi:20}, which is likely the cause of suppressed dipole modes in red giant stars\cite{stello:16}. Follow-up work showed that magnetic fields greater than $B \sim 0.1$ MG are sufficient to suppress g modes of typical periods in ZZ Ceti stars\cite{cantiello:16}, so magnetic suppression is a possibility for ZTF J1901+1458.

For these reasons, we believe the variability of ZTF J1901+1458 is most likely to be caused by rotation rather than pulsations, though we cannot rule out either mechanism. Follow-up observations can place more stringent limits on the presence of other non-harmonic periodicities which would be expected in the pulsation hypothesis. Finally, even if pulsations are the source of variability in ZTF J1901+1458, it would further enhance the extraordinary nature of this star by making it the most massive pulsating white dwarf and the only known magnetic pulsating white dwarf.

\end{methods}

\begin{addendum}
     \item[Data Availability] Upon request, I.C. will provide the reduced photometric lightcurves and spectroscopic data, and available ZTF data for the object. The spectroscopic data and photometric lightcurves are also available in the GitHub repository \texttt{https://github.com/ilac/ZTF-J1901-1458}, while ZTF data is accessible in the ZTF database. The astrometric and photometric data are already in the public domain, and they are readily accessible in the Gaia and Pan-STARSS catalogues and in the Swift database.
 
 \item[Code availability] We used the  \texttt{pyphot} package (\texttt{https://mfouesneau.github.io/docs/pyphot/}) and the \texttt{corner.py} package\cite{corner}. The LRIS spectra were reduced using the \texttt{Lpipe} pipeline\cite{2019PASP..131h4503P}. Upon request, I.C. will provide the code used to analyse the spectroscopic and photometric data.
 
\end{addendum}

\let\oldthebibliography=\thebibliography
\let\oldendthebibliography=\endthebibliography
\renewenvironment{thebibliography}[1]{%
     \oldthebibliography{#1}%
     \setcounter{enumiv}{ 32 }%
}{\oldendthebibliography}


\begin{addendum}
    \item The authors thank Shing-Chi Leung and Sterl Phinney for insightful discussions, and Nicole Reindl and Mukremin Kilic for the very useful feedback. I.C. is a Sherman Fairchild Fellow at Caltech and thanks the Burke Institute for supporting her research. J.F. is thankful for support through an Innovator Grant from The Rose Hills Foundation, and the Sloan Foundation through grant FG-2018-10515. K.B.B thanks the National Aeronautics and Space Administration and the Heising Simons Foundation for supporting his research. J.S. is supported by the A.F. Morrison Fellowship in Lick Observatory and by the National Science Foundation through grant ACI-1663688. This work was supported by the Natural Sciences and Engineering Research Council of Canada. This work is based on observations obtained with the Samuel Oschin Telescope 48-inch and the 60-inch Telescope at the Palomar Observatory as part of the Zwicky Transient Facility project. ZTF is supported by the National Science Foundation under Grant No. AST-1440341 and a collaboration including Caltech, IPAC, the Weizmann Institute for Science, the Oskar Klein Center at Stockholm University, the University of Maryland, the University of Washington (UW), Deutsches Elektronen-Synchrotron and Humboldt University, Los Alamos National Laboratories, the TANGO Consortium of Taiwan, the University of Wisconsin at Milwaukee, and Lawrence Berkeley National Laboratories. Operations are conducted by Caltech Optical Observatories, IPAC, and UW. Some of the data presented herein were obtained at the W.M. Keck Observatory, which is operated as a scientific partnership among the California Institute of Technology, the University of California and the National Aeronautics and Space Administration. This work has made use of data from the European Space Agency (ESA) mission {\it Gaia} \\(https://www.cosmos.esa.int/gaia), processed by the {\it Gaia} Data Processing and Analysis Consortium (DPAC, https://www.cosmos.esa.int/web/gaia/dpac/consortium). Funding for the DPAC has been provided by national institutions, in particular the institutions participating in the {\it Gaia} Multilateral Agreement. The Pan-STARRS1 Surveys (PS1) and the PS1 public science archive have been made possible through contributions by the Institute for Astronomy, the University of Hawaii, the Pan-STARRS Project Office, the Max-Planck Society and its participating institutes, the Max Planck Institute for Astronomy, Heidelberg and the Max Planck Institute for Extraterrestrial Physics, Garching, The Johns Hopkins University, Durham University, the University of Edinburgh, the Queen's University Belfast, the Harvard-Smithsonian Center for Astrophysics, the Las Cumbres Observatory Global Telescope Network Incorporated, the National Central University of Taiwan, the Space Telescope Science Institute, the National Aeronautics and Space Administration under Grant No. NNX08AR22G issued through the Planetary Science Division of the NASA Science Mission Directorate, the National Science Foundation Grant No. AST-1238877, the University of Maryland, Eotvos Lorand University (ELTE), the Los Alamos National Laboratory, and the Gordon and Betty Moore Foundation.
 
 \item[Author Contributions] I.C. reduced the UV data, conducted the spectral and photometric analysis, identified the magnetic field and is the primary author of the manuscript. K.B.B. performed the period search on ZTF data and reduced the optical data. I.C. and J.H. conducted the mass-radius analysis. I.C., K.B.B., J.F., J.H., S.R.K., T.A.P., H.B.R. and J.S. contributed to the physical interpretation of the object. J.S. constructed preliminary MESA models for the object. I.A., A.D., D.A.D., A.A.M., F.J.M., R.S. and M.T.S. contributed to the implementation of ZTF. G.H. is a co-PI of ZTF Mid-Scale Innovations Program (MSIP). M.J.G. is the project scientist, E.C.B. is the survey scientist, T.A.P. is the co-PI and S.R.K. is the PI of ZTF.
 
 \item[Competing Interests] The authors declare that they have no
competing financial interests.

 \item[Correspondence] Correspondence and requests for materials
should be addressed to I.C.~(email: ilariac@caltech.edu).
\end{addendum}

\renewcommand{\tablename}{Extended Data Table}
\setcounter{table}{0} 
\renewcommand{\figurename}{Extended Data Figure}
\setcounter{figure}{0} 

\begin{table}
\centering
 \caption{\textbf{Photometric data for ZTF~J1901+1458.}}
\medskip
\begin{tabular}{cccc}
\hline
 \makecell{$\rm{PS1-}$$z$} & \makecell{$\rm{PS1-}$$i$} & \makecell{$\rm{PS1-}$$r$}& \makecell{$\rm{PS1-}$$g$}\\
 $16.595\pm0.006$ & $16.276\pm0.003$ & $15.977\pm0.003$ & $15.513\pm0.002$ \\ 
\hline
\makecell{$\rm{UVOT-}$$U$} & \makecell{$\rm{UVOT-}$$UVW1$} & \makecell{$\rm{UVOT-}$$UVM2$} & \makecell{$\rm{UVOT-}$$UVW2$}  \\
$14.94\pm0.06$ & $14.42\pm0.06$ & $14.40\pm0.06$ & $14.38\pm0.06$  \\
\hline
\end{tabular}
\label{tab:photometry}
\end{table}

\begin{table}
\centering
 \caption{\textbf{Identified Balmer transitions}. The numbers correspond to the identified transitions in Fig.~\ref{fig:sp} and the wavelengths of the transitions at 800~MG are from\cite{1994asmf.book.....R}. For 7 and 11, we also list the wavelength for 600~MG (left).}
\medskip
\begin{tabular}{c|lccc}
\hline
& \makecell{$\rm{line}$} &\makecell{$nlm\,\rightarrow\,n'l'm'$} &\makecell{$NM\mu\,\rightarrow\,N'M'\mu'$} & \makecell{$\lambda~\rm{[\AA]}$} \\
\hline
1 & $\rm{Balmer}~\rm{H}\gamma$ & $2s_0 \,\rightarrow\, 5p^\prime_{0}$ & $0\;0\;2 \,\rightarrow\,0\;0\;11$ & $3344.23$ \\ 
2 & $ \rm{Balmer}~\rm{H}\gamma$ & $2s_0 \,\rightarrow \, 5p^\prime_{-1}$ & $0\;0\;2 \,\rightarrow\,0\;\rm{-}1\;10$ & $3370.41$ \\
3 & $ \rm{Balmer}~\rm{H}\gamma$ & $2s_0 \,\rightarrow \, 5f^\prime_{0}$ & $0\;0\;2 \,\rightarrow\,0\;0\;9$ & $3464.11$ \\ 
4 & $ \rm{Balmer}~\rm{H}\beta$ & $2s_0 \,\rightarrow \, 4p^\prime_{0}$ & $0\;0\;2 \,\rightarrow\,0\;0\;7$ & $3694.56$\\ 
5 & $ \rm{Balmer}~\rm{H}\beta$ & $2s_0 \,\rightarrow \, 4p^\prime_{-1}$ & $0\;0\;2 \,\rightarrow\,0\;\rm{-}1\;6$ & $3805.82$ \\ 
6 & $ \rm{Balmer}~\rm{H}\beta$ & $2s_0 \,\rightarrow \, 4f^\prime_{0}$ & $0\;0\;2 \,\rightarrow\,0\;0\;5$ & $4249.86$ \\ 
7 & $ \rm{Balmer}~\rm{H}\beta$ & $2s_0 \,\rightarrow \, 4f^\prime_{-1}$ & $0\;0\;2 \,\rightarrow\,0\;\rm{-}1\;4$ & $4517.97,\,4590.9$ \\ 
8 & $ \rm{Balmer}~\rm{H}\alpha$ & $2p_{-1} \,\rightarrow \, 3d_{-2}$ & $0\;\rm{-}1\;0 \,\rightarrow\,0\;\rm{-}2\;0$ & $5883.63$  \\
9 & $ \rm{Balmer}~\rm{H}\alpha$ & $2s_0 \,\rightarrow \, 3p_{0}$ & $0\;0\;2 \,\rightarrow\,0\;0\;3$ & $6615.06$ \\
10 & $ \rm{Balmer}~\rm{H}\alpha$ & $2p_0 \,\rightarrow \, 3d_{-1}$ & $0\;0\;2 \,\rightarrow\,0\;\rm{-}1\;1$ & $8525.69$  \\
11 & $ \rm{Balmer}~\rm{H}\alpha$ & $2s_0 \,\rightarrow \, 3p_{-1}$ & $0\;0\;1 \,\rightarrow\,0\;\rm{-}1\;2$ & $8537.76,\,9542.62$  \\
\hline
\end{tabular}
\label{tab:transitions}
\end{table}

\renewcommand{\figurename}{Extended Data Figure}
\setcounter{figure}{0}

\begin{figure}
    \centering
    \includegraphics[width=\textwidth]{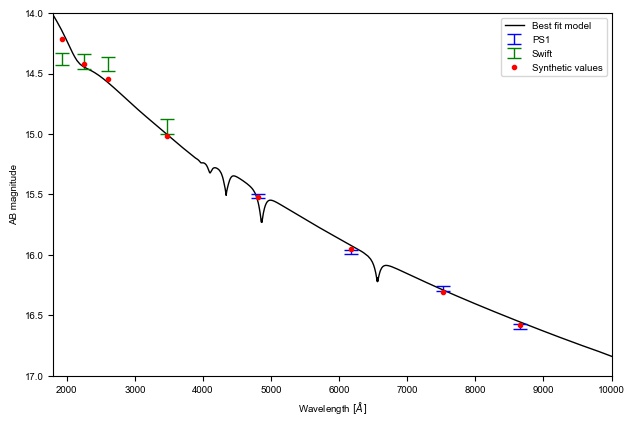}
    \caption{{\bf Photometric fit.} The blue solid line shows the best-fitting model spectrum, fitted to Pan-STARRS and Swift photometry to determine $T_{\rm{eff}}$, $R_*$ and $E(B-V)$. The synthetic photometric values (obtained from the black line) are shown in red, while the Swift values are shown in green with 1 $\sigma$ error bars and the Pan-STARRS values in blue with the error that we chose to account for the photometric variability (0.02 mag).} 
    \label{fig:phot}
\end{figure}

\begin{figure}
    \centering
    \includegraphics[width=0.8\textwidth]{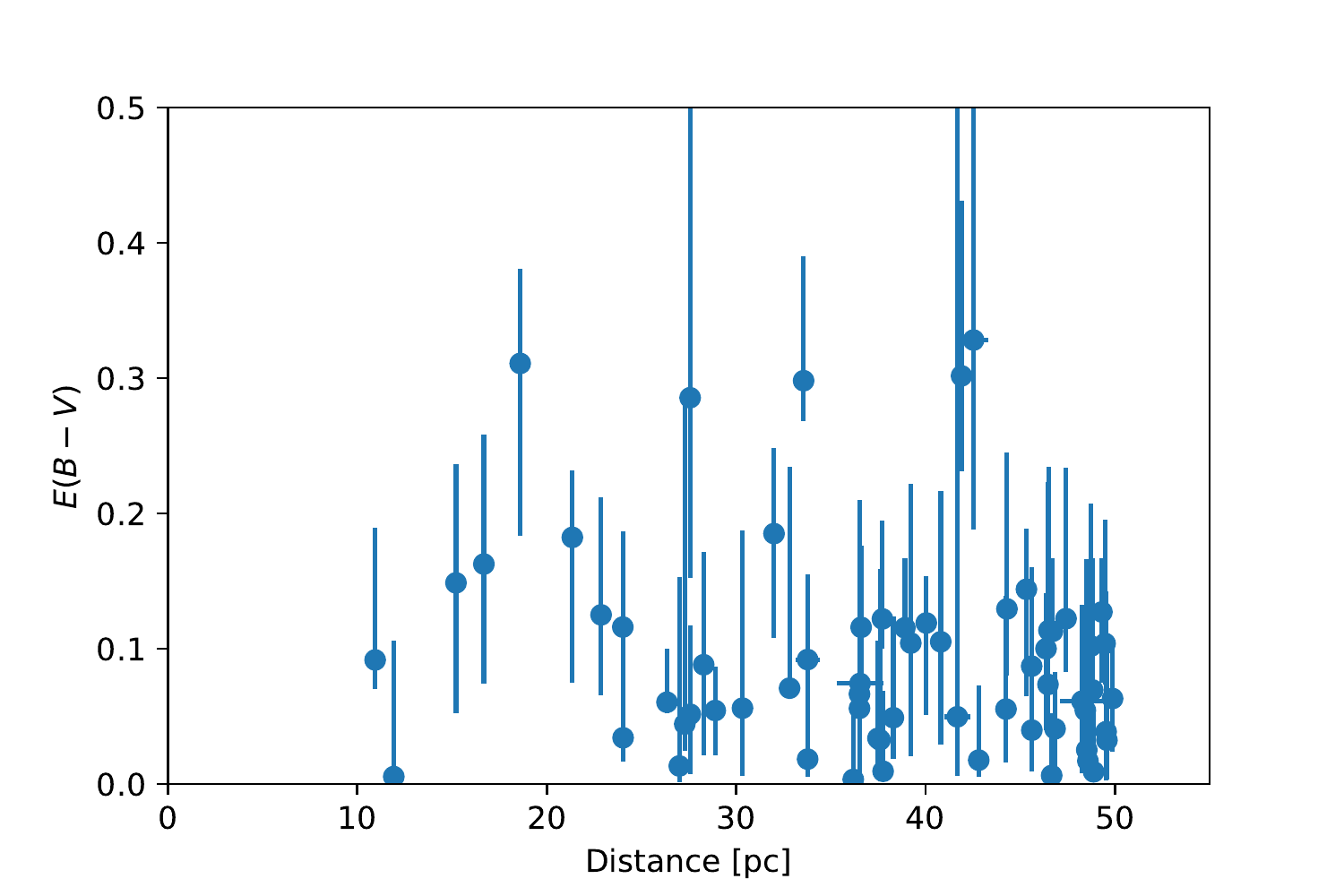}
    \caption{{\bf Gaia $E(B-V)$ of the closest stars. }The plot shows the extinction $A_G$ as  measured by Gaia of the stars within 5 degrees of ZTF 1901+1458 as a function of distance and converted to $E(B-V)$ assuming a reddening law with $R_V=3.1$. We use the average reddening or the closest stars as a prior for the fitting. The error bars show 1$\sigma$ errors.} 
    \label{fig:ebmv}
\end{figure}

\begin{figure}
    \centering
    \includegraphics[width=0.495\textwidth]{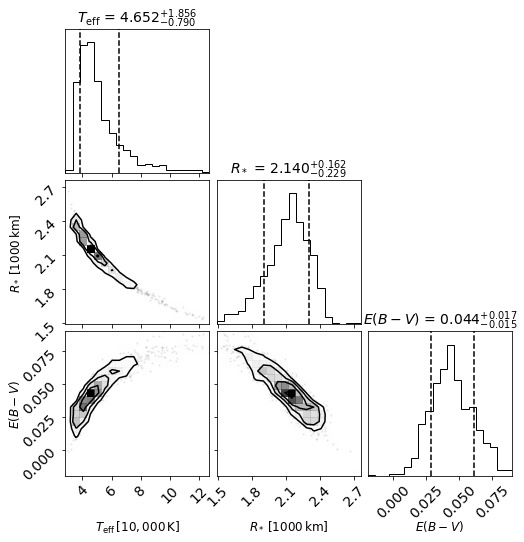}
    \includegraphics[width=0.495\textwidth]{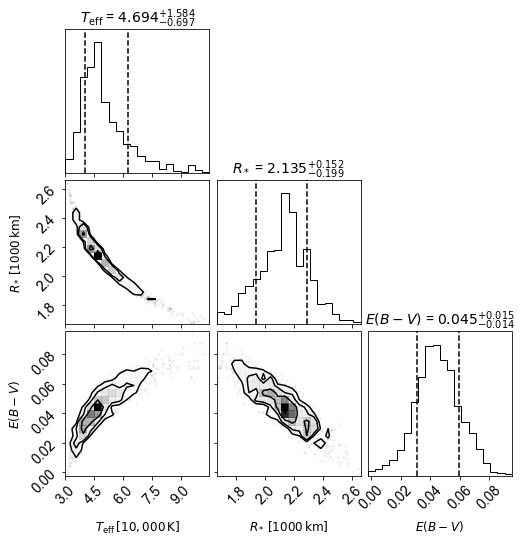}
    \caption{{\bf Corner plots.} Corner plots for the photometric fitting  \textbf{Left}: results for the model atmospheres of Tremblay et al. 2011\cite{2011ApJ...730..128T}, \textbf{Right}: results for the model atmospheres of Bohlin et al 2020\cite{2020AJ....160...21B}.} 
    \label{fig:corner}
\end{figure}

\begin{figure}
    \centering
    \includegraphics[width=0.9\textwidth]{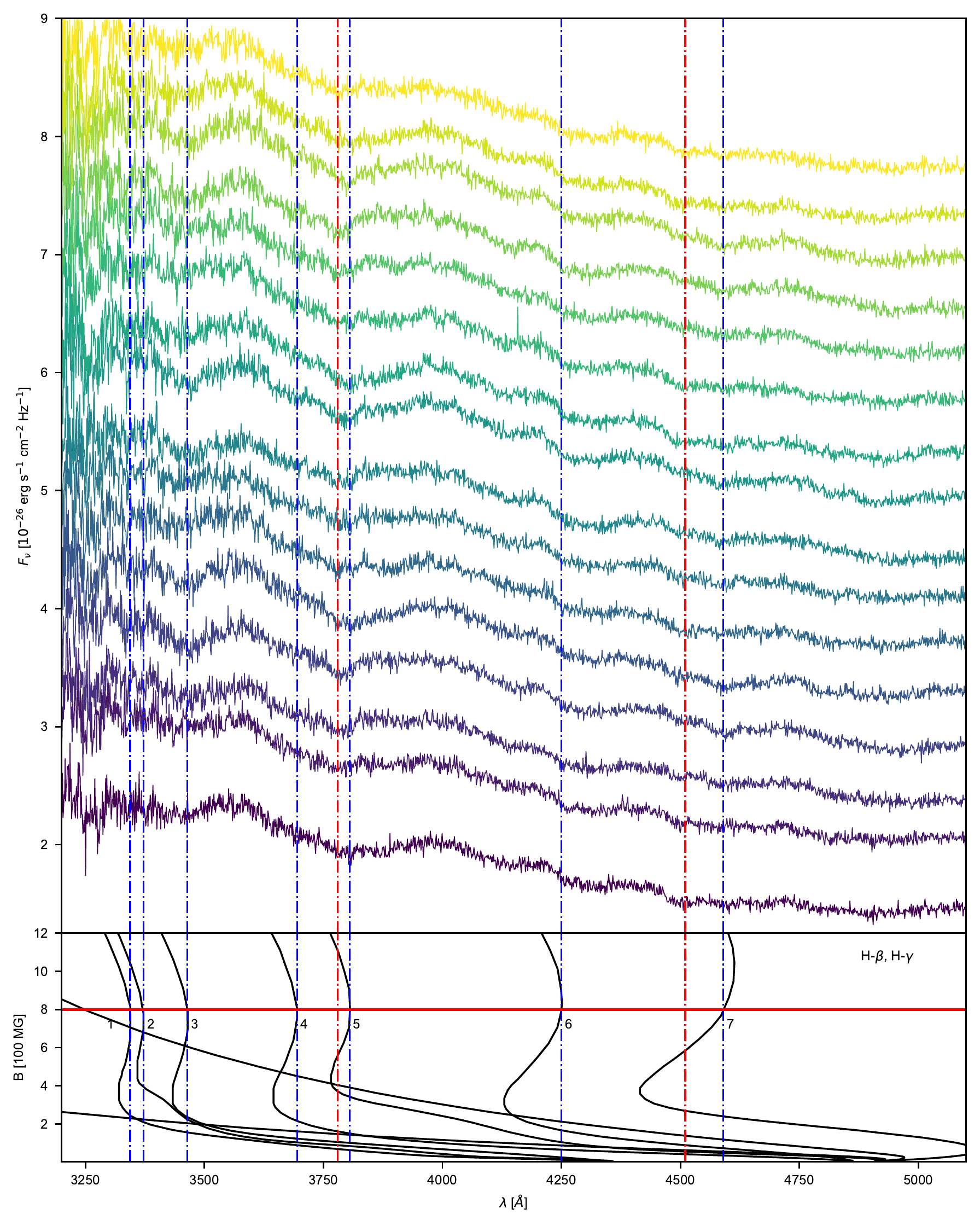}
    \caption{{\bf Phase-resolved spectra, blue side.} The LRIS phase-resolved spectra of ZTF J1901+1458 in the blue side. Some small variations can be observed in the spectral features with phase, especially in features at $\sim4600$~\AA\ and at $\sim3800$~\AA.} 
    \label{fig:pr1}
\end{figure}

\begin{figure}
    \centering
    \includegraphics[width=0.9\textwidth]{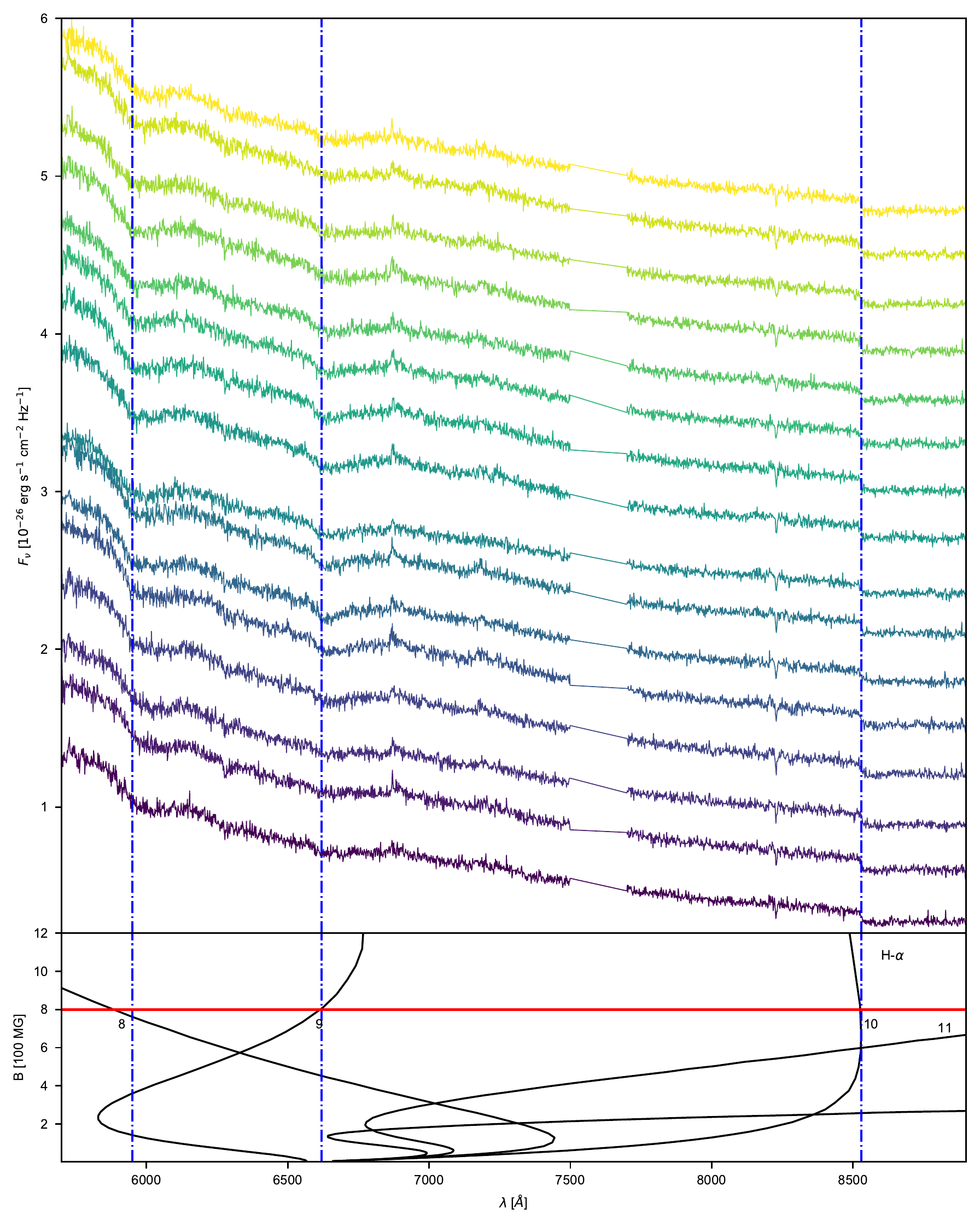}
    \caption{{\bf Phase-resolved spectra, red side.} The LRIS phase-resolved spectra of ZTF J1901+1458 in the red side. Some small variations can be observed in the spectral features with phase, in particular the feature at $\sim6620$~\AA\ becomes broader and narrower with phase. } 
    \label{fig:pr2}
\end{figure}

\end{document}